\documentclass[12pt,aps,prb,preprint]{revtex4}   

\usepackage{amsmath}    
\usepackage{amssymb}
\usepackage{graphicx}   

\usepackage[latin1]{inputenc}     
\usepackage[T1]{fontenc}

\usepackage{float}   

\newtheorem{Theoreme}{THEOREME}
\newtheorem{Definition}{Définition}

\draft

\begin{document}

\title{From interpretation of the three classical mechanics actions to the wave function in quantum mechanics }
\author{Michel Gondran}
 \affiliation{University Paris Dauphine, 75 016 Paris.}
 \email{michel.gondran@polytechnique.org}   
\author{Alexandre Gondran}
 \affiliation{Ecole Nationale de l'Aviation Civile, 31000 Toulouse, France}

\begin{abstract}
First, we show that there exists in classical mechanics three
actions corresponding to different boundary conditions: two
well-known actions, the Euler-Lagrange classical action
$S_{cl}(\textbf{x},t;\textbf{x}_0)$, which links the initial
position $\textbf{x}_0$ and its position $\textbf{x}$ at time t,
the Hamilton-Jacobi action $S(\textbf{x},t)$, which links a family
of particles of initial action $S_{0}(\textbf{x})$ to their
various positions $\textbf{x}$ at time t, and a new action, the
deterministic action $S(\textbf{x},t;\textbf{x}_0,\textbf{x}_0)$,
which links a particle in initial position $\textbf{x}_0$ and
initial velocity $\textbf{v}_0$ to its position $\textbf{x}$ at
time t. Mathematically, the Euler-Lagrange action can be
considered as the elementary solution to the Hamilton-Jacobi
equation in a new branch of non-linear mathematics, the Minplus
analysis. We study, in the semi-classical approximation, the
convergence of the quantum density and the quantum action,
solutions to the Madelung equations, when the Planck constant h
tends to 0. We find two different solutions which depend on the
initial density. In the first case, where the initial quantum
density is a classical density $\rho_{0}(\mathbf{x})$, the quantum
density and the quantum action converge to a classical action and
a classical density which satisfy the statistical Hamilton-Jacobi
equations. These are the equations of a set of classical particles
whose initial positions are known only by the density
$\rho_{0}(\mathbf{x})$. In the second case where initial density
converges to a Dirac density, the density converges to the Dirac
function and the quantum action converges to a deterministic
action. Therefore we introduce into classical mechanics
non-discerned particles, which satisfy the statistical
Hamilton-Jacobi-equations and explain the Gibbs paradox, and
discerned particles, which satisfy the deterministic
Hamilton-Jacobi equations. When the semi-classical approximation
is not valid, we conclude that the Schrödinger equation cannot
give a deterministic interpretation and the statistical Born
interpretation is the only valid one. Finally, we propose an
interpretation of the Schr\"odinger wave function that depends on
the initial conditions (preparation). This double interpretation
seems to be the interpretation of Louis de Broglie's "double
solution" idea.
\end{abstract}

\maketitle

\section{Introduction}

The aim of this paper is to show how the interpretation of the
wave function in quantum mechanics can be deduced from the
interpretation of the action in classical mechanics and from the
study of the convergence QM-CM when the Planck constant tends to
0. First, in section 2, we show that there exist in classical
mechanics three actions corresponding to different boundary
conditions: two well-known actions, the Euler-Lagrange action and
the Hamilton-Jacobi action, and a new action, the deterministic
action. We introduce these three actions and present the
fundamental relation between the Hamilton-Jacobi and
Euler-Lagrange actions. Second, in section 3, we present a new
branch of non-linear mathematics, the Minplus analysis that we
have developed following Maslov. In this new analysis, the
Hamilton-Jacobi equation can be considered as linear. Third, in
section 4, we present in the semiclassical case approximation, the
QM-CM convergence when the Planck constant tends to 0. It is
necessary to introduce two cases: the statistical semi-classical
case and the deterministic semi-classical case. Fourth, in section
5, we discuss the case where the semi-classical approximation is
not valid. Finally, we propose a realistic interpretation of
quantum mechanics, which is a synthesis of the three
interpretations of the founding fathers of quantum mechanics at
the Solvay congress in 1927: the de Broglie interpretation, the
Schrödinger interpretation and the Copenhagen interpretation.

\section{The three classical mechanics actions}

Let us consider a system evolving from the position
$\textbf{x}_{0}$ at initial time $t_0=0$ to the position
$\textbf{x}$ at time t where the variable of control \textbf{u}(s)
is the velocity:
\begin{equation}\label{eq:evolution}
\frac{d \textbf{x}\left( s\right) }{ds}=\mathbf{u}(s)~~~~
for~~~~s\in \left[ 0,t\right]
\end{equation}
\begin{equation}\label{eq:condinitiales}
\textbf{x}(0) =\mathbf{x}_{0},~~~~~~\textbf{x}(t) =\mathbf{x}.
\end{equation}

If $L(\textbf{x},\dot{\textbf{x}},t)$ is the Lagrangian of the
system, when the two positions $\textbf{x}_0$ and $\textbf{x}$ are
given, \emph{the Euler-Lagrange action} $S_{cl}(\mathbf{x},t;
\textbf{x}_0) $ is the function defined by:
\begin{equation}\label{eq:defactioncondit}
S_{cl}(\mathbf{x},t;\textbf{x}_0)=\min_{\mathbf{u}\left(
s\right),0 \leq s\leq t}\left\{ \int_{0}^{t}L(\textbf{x}(s),%
\mathbf{u}(s),s)ds\right\},
\end{equation}
where the minimum (or more generally an extremum) is taken on the
controls $\mathbf{u}(s)$, $s\in$ $\left[ 0,t\right]$, with the
state $\textbf{x}(s)$ given by the equations
(\ref{eq:evolution})(\ref{eq:condinitiales}). The solution
$(\widetilde{\textbf{u}}(s), \widetilde{\textbf{x}}(s))$ of
(\ref{eq:defactioncondit}) satisfies the Euler-Lagrange equations
on the interval $[0,t] $:
\begin{equation}\label{eq:EulerLagrange1}
\frac{d}{ds}\frac{\partial L}{\partial
\dot{\textbf{x}}}(\textbf{x}(s),\dot{\textbf{x}}(s),s)-
\frac{\partial L}{\partial
\textbf{x}}(\textbf{x}(s),\dot{\textbf{x}}(s),s)=0~~~~~~~~~~(0\leq
s \leq t)
\end{equation}
\begin{equation}\label{eq:EulerLagrange12}
\textbf{x}(0) =\mathbf{x}_{0},~~~~~~\textbf{x}(t) =\mathbf{x}.
\end{equation}

If $ L(\mathbf{x},\dot{\textbf{x}},t)= \frac{1}{2}m
\dot{\textbf{x}}^2 + \textbf{K}. \textbf{x}$, then the
Euler-Lagrange action is $ S_{cl}( \mathbf{x},t; \textbf{x}_0)= m
\frac{(\textbf{x}-\textbf{x}_0)^2}{2 t}+ \frac{K .(\textbf{x} +
\textbf{x}_0)}{2}t - \frac{K^2}{24 m}t^3$ and the initial velocity
is given by $ \textbf{v}_0= \dot{\textbf{x}}(o)= -
\frac{1}{m}\frac{\partial S_{cl}}{\partial \textbf{x}_0}(
\mathbf{x},t; \textbf{x}_0)=\frac{\textbf{x}
-\textbf{x}_0}{t}-\frac{K t}{2 m}.$

Let us now consider that an initial action $S_0(\textbf{x})$ is
given, then \emph{the Hamilton-Jacobi action } $S(\mathbf{x},t) $
is the function defined by:
\begin{equation}\label{eq:defactionHJ}
S(\mathbf{x},t)=\min_{\textbf{x}_0;\mathbf{u}\left( s\right),0
\leq s\leq t }\left\{ S_{0}\left( \mathbf{x}_{0}\right) +\int_{0}^{t}L(\textbf{x}(s),%
\mathbf{u}(s),s)ds\right\}
\end{equation}
where the minimum is taken on all initial positions
$\textbf{x}_0$, on the controls $\mathbf{u}(s)$, $s\in$ $\left[
0,t\right]$, with the state $\textbf{x}(s)$ given by the equations
(\ref{eq:evolution})(\ref{eq:condinitiales}). Because the term
$S_0(\textbf{x}_0)$ has no effect in equation
(\ref{eq:defactionHJ}) for the minimization on the control
$\textbf{u}(s)$, we deduce the important relation between the
Hamilton-Jacobi action and Euler-Lagrange action:
\begin{equation}\label{eq:ELHJ}
S(\mathbf{x},t)=\min_{\textbf{x}_0} ( S_{0}\left(
\mathbf{x}_{0}\right) + S_{cl}(\textbf{x},t;\textbf{x}_0) ).
\end{equation}
This equation is similar to the Hopf-Lax or Lax-Oleinik
formula~\cite{Evans}.

If $ L(\mathbf{x},\dot{\textbf{x}},t)= \frac{1}{2}m
\dot{\textbf{x}}^2 + \textbf{K}. \textbf{x}$ with the initial
action $ S_0(\textbf{x})= m \textbf{v}_0 \cdot \textbf{x} $, then
the Hamilton-Jacobi action is equal to $
 S\left( \mathbf{x},t\right)=m \textbf{v}_0 \cdot
\textbf{x} - \frac{1}{2} m \textbf{v}_0^2 t +\textbf{K}.\textbf{x}
t - \frac{1}{2} \textbf{K}.\textbf{v}_{0} t^{2} -
\frac{\textbf{K}^2 t^3}{6 m}.$

For  a non-relativistic particle with the Lagrangian
$L(\mathbf{x},\mathbf{\dot{x}},t)= \frac{1}{2}m \mathbf{\dot{x}}^2
- V(\textbf{x},t)$, we obtain the well-known result: \emph{The
velocity of a non-relativistic classical particle in a potential
field is given for each point} $ \left( \mathbf{x,}t\right)$
\emph{by}:
\begin{equation}\label{eq:eqvitesse}
\mathbf{v}\left( \mathbf{x,}t\right) =\frac{\mathbf{\nabla }S\left( \mathbf{%
x,}t\right) }{m}
\end{equation}
\textit{where} $S\left( \mathbf{x,}t\right) $\textit{\ is the
Hamilton-Jacobi action, a solution to the Hamilton-Jacobi
equations:}
\begin{equation}\label{eq:HJ}
\frac{\partial S(\textbf{x},t)}{\partial t}+\frac{1}{2m}(\nabla
S(\textbf{x},t) )^{2}+V(\textbf{x},t)=0
\end{equation}
\begin{equation}\label{eq:condinitialHJ}
S(\textbf{x},0)=S_{0}(\textbf{x}).
\end{equation}

The Hamilton-Jacobi action corresponds to a velocity field $\mathbf{v}\left( \mathbf{x,}t\right) =\frac{\mathbf{\nabla }S\left( \mathbf{%
x,}t\right) }{m}$. The Hamilton-Jacobi action $S(\mathbf{x},t)$
does not solve only a given problem with a single initial
condition $\left( \mathbf{x}_{0}, \frac{\mathbf{\nabla
}S_{0}\left( \mathbf{x}_{0}\right) }{m}\right) $, but a set of
problems with an infinity of initial conditions $\left( \mathbf{y},%
\frac{\mathbf{\nabla }S_{0}\left( \mathbf{y}\right) }{m}\right)$.
It is the problem solved by Nature with the principle of least
action.

In the absence of an initial velocity field as in the
Hamilton-Jacobi action, the Euler-Lagrange action answers a
problem posed by the observer, and not by Nature: "If we see that
a particle in $\textbf{x}_0$ at the initial time arrives in
$\textbf{x}$ at time t, what was its initial velocity $\textbf
{v}_0$?"

Let us now consider that we know the initial conditions
($\textbf{x}_0$, $\textbf {v}_0$) and the Lagrangian of the
system. If $\xi(t)$ is the classical trajectory in the field
$V(\textbf{x},t)$ of the particle with the initial conditions
$\left( \mathbf{x}_{0}\mathbf{,v}_{0}\right)$, then we define
\emph{the deterministic action} $S( \mathbf{x},t;\textbf{x}_0,
\textbf{v}_0)$ by the equation:
\begin{equation}\label{eq:soleqHJponctuelle}
S( \mathbf{x},t;\textbf{x}_0, \textbf{v}_0)= m \frac{d \xi(t)}{dt}
\cdot \textbf{x} + g(t)
\end{equation}
where $g(t)= -\int^{t}_{0}{\frac{1}{2}m (\frac{d \xi(s)}{ds})^2 +
V(\xi(s)) + m \frac{d^2 \xi(s)}{ds^2} \cdot \xi(s)} ds$.

\begin{Theoreme}\label{th:actionponctuelle} The deterministic action is a solution
to the deterministic Hamilton-Jacobi equations:
\begin{equation}\label{eq:statHJponctuelle1b}
\frac{\partial S(\textbf{x},t;\textbf{x}_0,\textbf{v}_0)
}{\partial t}|_{\textbf{x}=\xi(t)}+\frac{1}{2m}(\nabla
S(\textbf{x},t;\textbf{x}_0, \textbf{v}_0)
)^{2}|_{\textbf{x}=\xi(t)}+V(\textbf{x})|_{\textbf{x}=\xi(t)}=0
\end{equation}
\begin{equation}\label{eq:statHJponctuelle1c}
\frac{d\xi(t)}{dt}=\frac{\nabla S(\textbf{x},t;\textbf{x}_0,
\textbf{v}_0)}{m}|_{\textbf{x}=\xi(t)}
\end{equation}
\begin{equation}\label{eq:statHJponctuelle1d}
S(\textbf{x},0;\textbf{x}_0,\textbf{v}_0)= m \textbf{v}_0
\textbf{x} ~~~and~~~\xi(0)=\textbf{x}_0.
\end{equation}
\end{Theoreme}
The deterministic action satisfies the Hamilton-Jacobi equations
only along the classical trajectory $\xi(t)$. It is the action
introduced by Rybakov~\cite{Rybakov} for a soliton. We will
interpret these equations in section 4 when we will study the
QM-CM convergence.

\section{Interpretation of the Euler-Lagrange in Minplus analysis}

There exists a new branch of mathematics, the Minplus analysis,
which studies nonlinear problems through a linear approach, cf.
Maslov~\cite{Maslov,Maslov2} and
Gondran~\cite{Gondran_1996,GondranMinoux}. The idea is to
substitute the usual scalar product $\int_{X} f(x) g(x) dx$ with
the Minplus scalar product:
\begin{equation}
    (f,g) =\inf_{x\in X}\left\{ f(x)+g(x) \right\}
\end{equation}
In the scalar product we replace the field of the real number $(
R,+,\times )$ with the algebraic structure \textit{Minplus} $(
R\cup \{ +\,\infty \} ,\min ,+)$, i.e. the set of real numbers
(with the element infinity $\{ +\infty \}$) endowed with the
operation Min (minimum of two reals), which remplaces the usual
addition, and with the operation + (sum of two reals), which
remplaces the usual multiplication. The element $\{+\,\infty \}$
corresponds to the neutral element for the operation Min, Min$( \{
+\infty \} ,a) =a$ $\forall a\in R$. This approach bears a close
similarity to \emph{the theory of distributions for the nonlinear
case}; here, the operator is "linear" and continuous with respect
to the Minplus structure,
though \emph{nonlinear} with respect to the classical structure $%
\left( R,+,\times \right)$. In this Minplus structure, the
Hamilton-Jacobi equation is linear, because if $S_1(\textbf{x},t)$
and $S_2(\textbf{x},t)$ are solutions to (\ref{eq:HJ}), then
$\min\{\lambda + S_1(\textbf{x},t), \mu + S_2(\textbf{x},t)\}$ is
also a solution to the Hamilton-Jacobi equation (\ref{eq:HJ}).

The analog to the Dirac distribution $\delta(\textbf{x})$ in
Minplus analysis is the nonlinear distribution
$\delta_{\min}(\textbf{x})=\{ 0~if~\textbf{x}=\textbf{0}, +\infty~
if~not\}$. With this nonlinear Dirac distribution, we can define
elementary solutions as in classical distribution theory. In
particular, we have:

\emph{The classical Euler-Lagrange action
$S_{cl}(\textbf{x},t;\textbf{x}_0)$ is the elementary solution to
the Hamilton-Jacobi equations
(\ref{eq:HJ})(\ref{eq:condinitialHJ}) in the Minplus analysis with
the initial condition}
\begin{equation}
S(\textbf{x},0)= \delta_{\min}(\textbf{x}- \textbf{x}_0)= \{
0~~if~~\textbf{x}=\textbf{x}_0,~+\infty~~if~not \}.
\end{equation}
The Hamilton-Jacobi action $S(\textbf{x},t)$ is then given by the
Minplus integral:
\begin{equation}
S(\textbf{x},t)=\inf_{\textbf{x}_0} \{ S_0(\textbf{x}_0)
 + S_{cl}(\textbf{x},t;\textbf{x}_0)\}
\end{equation}
in analogy with the solution to the heat transfer equation given
by the classical integral:
\begin{equation}
        S(x,t)=\int S_{0}( x_0) \frac{1}{2\sqrt{\pi t}} e^{-\frac{
        \left( x-x_0\right) ^{2}}{4t}}dx_0 .
\end{equation}

In this Minplus analysis, the Legendre-Fenchel transform is the
analog to the Fourier transform. This transform is known to have
many applications in physics: it sets the correspondence between
the Lagrangian and the Hamiltonian of a physical system; it sets
the correspondence between microscopic and macroscopic models; it
is also at the basis of multifractal analysis relevant to modeling
turbulence in fluid mechanics \cite{GondranMinoux}.

\section{The two limits of the Schr\"odinger equation in the semi-classical approximation}

Let us consider the wave function solution to the Schr\"odinger
equation $\Psi(\textbf{x},t)$:
\begin{equation}\label{eq:schrodinger1}
i\hbar \frac{\partial \Psi }{\partial t}=\mathcal{-}\frac{\hbar ^{2}}{2m}%
\triangle \Psi +V(\mathbf{x},t)\Psi
\end{equation}
\begin{equation}\label{eq:schrodinger2}
\Psi (\mathbf{x},0)=\Psi_{0}(\mathbf{x}).
\end{equation}
With the variable change $ \Psi
(\mathbf{x},t)=\sqrt{\rho^{\hbar}(\mathbf{x},t)} \exp(i
\frac{S^{\hbar}(\textbf{x},t)}{\hbar})$, the Schr\"odinger
equation can be decomposed into Madelung
equations~\cite{Madelung_1926} (1926):
\begin{equation}\label{eq:Madelung1}
\frac{\partial S^{\hbar}(\mathbf{x},t)}{\partial t}+\frac{1}{2m}
(\nabla S^{\hbar}(\mathbf{x},t))^2 +
V(\mathbf{x},t)-\frac{\hbar^2}{2m}\frac{\triangle
\sqrt{\rho^{\hbar}(\mathbf{x},t)}}{\sqrt{\rho^{\hbar}(\mathbf{x},t)}}=0
\end{equation}
\begin{equation}\label{eq:Madelung2}
\frac{\partial \rho^{\hbar}(\mathbf{x},t)}{\partial t}+ div
(\rho^{\hbar}(\mathbf{x},t) \frac{\nabla
S^{\hbar}(\mathbf{x},t)}{m})=0
\end{equation}
with initial conditions
\begin{equation}\label{eq:Madelung3}
\rho^{\hbar}(\mathbf{x},0)=\rho^{\hbar}_{0}(\mathbf{x}) \qquad and
\qquad S^{\hbar}(\mathbf{x},0)=S^{\hbar}_{0}(\mathbf{x}) .
\end{equation}
We consider two cases \textit{depending on the preparation of the
particles}~\cite{Gondran2011,Gondran2012}.
\begin{Definition}\label{defdensiteinitstat}- The \textbf{statistical semi-classical case}
where

- the initial probability density $\rho^{\hbar}_{0}(\mathbf{x})$
and the initial action $S^{\hbar}_{0}(\mathbf{x})$ are regular
functions $\rho_{0}(\mathbf{x})$ and $S_{0}(\mathbf{x})$ not
depending on $\hbar$.

- the interaction with the potential field $V(\textbf{x},t)$ can
be described classically.
\end{Definition}
It is the case of a set of non-interacting particles all prepared
in the same way: a free particle beam in a linear potential, an
electronic or $C_{60}$ beam in the Young's slits diffraction, or
an atomic beam in the Stern and Gerlach experiment.
\begin{Definition}\label{defdensiteinitponct}- The \textbf{determinist semi-classical case}
where

- the initial probability density $\rho^{\hbar}_{0}(\mathbf{x})$
converges, when $\hbar\to 0$, to a Dirac distribution and the
initial action $S^{\hbar}_{0}(\mathbf{x})$ is a regular function
$S_{0}(\mathbf{x})$ not depending on $\hbar$.

- the interaction with the potential field $V(\textbf{x},t)$ can
be described classically.
\end{Definition}
This situation occurs when the wave packet corresponds to a
quasi-classical coherent state, introduced in 1926 by
Schr\"odinger~\cite{Schrodinger_26}. The field quantum theory and
the second quantification are built on these coherent
states~\cite{Glauber_65}. The existence for the hydrogen atom of a
localized wave packet whose motion is on the classical trajectory
(an old dream of Schr\"odinger's) was predicted in 1994 by
Bialynicki-Birula, Kalinski, Eberly, Buchleitner et
Delande~\cite{Bialynicki_1994, Delande_1995, Delande_2002}, and
discovered recently by Maeda and Gallagher~\cite{Gallagher} on
Rydberg atoms.

\begin{Theoreme}~\cite{Gondran2011,Gondran2012} For particles in the statistical semi-classical case,
the probability density $\rho^{\hbar}(\textbf{x},t)$ and the
action $S^{\hbar}(\textbf{x},t)$, solutions to the Madelung
equations
(\ref{eq:Madelung1})(\ref{eq:Madelung2})(\ref{eq:Madelung3}),
converge, when $\hbar\to 0$, to the classical density
$\rho(\textbf{x},t)$ and the classical action $S(\textbf{x},t)$,
solutions to the statistical Hamilton-Jacobi equations:
\begin{equation}\label{eq:statHJ1}
\frac{\partial S\left(\textbf{x},t\right) }{\partial
t}+\frac{1}{2m}(\nabla S(\textbf{x},t) )^{2}+V(\textbf{x},t)=0
\end{equation}
\begin{equation}\label{eq:statHJ3}
\frac{\partial \mathcal{\rho }\left(\textbf{x},t\right) }{\partial
t}+ div \left( \rho \left( \textbf{x},t\right) \frac{\nabla
S\left( \textbf{x},t\right) }{m}\right) =0\textit{ \ \ \ \ \ \ \
}\forall \left( \textbf{x},t\right)
\end{equation}
\begin{equation}\label{eq:statHJ4}
\rho(\mathbf{x},0)=\rho_{0}(\mathbf{x}) \qquad and \qquad
S(\textbf{x},0)=S_{0}(\textbf{x}).
\end{equation}
\end{Theoreme}

We give some indications on the demonstration of this theorem and
we propose its interpretation. Let us consider the case where the
wave function $\Psi(\textbf{x},t)$ at time t is written as a
function of the initial wave function $\Psi_{0}(\textbf{x})$ by
the Feynman paths integral formula \cite{Feynman_1965} (p. 58):
\begin{equation}\nonumber
\Psi(\textbf{x},t)= \int F(t,\hbar)
\exp(\frac{i}{\hbar}S_{cl}(\textbf{x},t;\textbf{x}_{0})
\Psi_{0}(\textbf{x}_{0})d\textbf{x}_0
\end{equation}
where $F(t,\hbar)$ is an independent function of $\textbf{x}$ and
of $\textbf{x}_{0}$ and where
$S_{cl}(\textbf{x},t;\textbf{x}_{0})$ is the classical action. In
the statistical semi-classical case, the wave function is written
$ \Psi(\textbf{x},t)= F(t,\hbar)\int\sqrt{\rho_0(\mathbf{x}_0)}
\exp(\frac{i}{\hbar}( S_0(\textbf{x}_0)+
S_{cl}(\textbf{x},t;\textbf{x}_{0})) d\textbf{x}_0$. The theorem
of the stationary phase shows that, if $\hbar$ tends towards 0, we
have $ \Psi(\textbf{x},t)\sim
\exp(\frac{i}{\hbar}min_{\textbf{x}_0}( S_0(\textbf{x}_0)+
S_{cl}(\textbf{x},t;\textbf{x}_{0}))$, that is to say that the
quantum action $S^{h}(\textbf{x},t)$ converges to the function
\begin{equation}\label{eq:solHJminplus}
S(\textbf{x},t)=min_{\textbf{x}_0}( S_0(\textbf{x}_0)+
S_{cl}(\textbf{x},t;\textbf{x}_{0}))
\end{equation}
which is the solution to the Hamilton-Jacobi equation
(\ref{eq:HJ}) with the initial condition (\ref{eq:condinitialHJ}).
Moreover, as the quantum density $\rho^{h}(\textbf{x},t)$
satisfies the continuity equation (\ref{eq:Madelung2}), we deduce,
since $S^{h}(\textbf{x},t)$ tends towards $S(\textbf{x},t)$, that
$\rho^{h}(x,t)$ converges to the classical density
$\rho(\textbf{x},t)$, which satisfies the continuity equation
(\ref{eq:statHJ3}). We obtain both announced convergences.

The statistical Hamilton-Jacobi equations correspond to a set of
independent classical particles, in a potential field
$V(\mathbf{x},t)$, and for which we only know at the initial time
the probability density $\rho _{0}\left( \mathbf{x}\right) $ and
the velocity $\mathbf{v(x)}=\frac{\nabla S_{0}(\mathbf{x},t)}{m}$.
\begin{Definition}\label{defNparticulesindiscernedenmc}- N identical particles,
prepared in the same way, with the same initial density $ \rho
_{0}\left( \textbf{x}\right)$, the same initial action
$S_0(\textbf{x})$, and evolving in the same potential
$V(\textbf{x},t)$ are called non-discerned.
\end{Definition}
We refer to these particles as non-discerned and not as
indistinguishable because, if their initial positions are known,
their trajectories will also be known. Nevertheless, when one
counts them, they will have the same properties as the
indistinguishable ones. Thus, if the initial density $\rho
_{0}\left( \textbf{x}\right)$ is given, and one randomly chooses
$N$ particles, the N! permutations are strictly equivalent and do
not correspond to the same configuration as for indistinguishable
particles.  This indistinguishability of classical particles
provides a very simple and natural explanation to the Gibbs
paradox.

In the statistical semi-classical case, the uncertainity about the
position of a quantum particle corresponds to an uncertainity
about the position of a classical particle, whose initial density
alone has been defined. \emph{In classical mechanics, this
uncertainity is removed by giving the initial position of the
particle. It would be illogical not to do the same in quantum
mechanics.} We assume that for \textit{the statistical
semi-classical case}, a quantum particle is not well described by
its wave function. One therefore needs  to add its initial
position and it follows that we introduce the so-called de
Broglie-Bohm trajectories \cite{deBroglie_1927,Bohm_52} with the
velocity $\textbf{v}^{\hbar}(\textbf{x},t) = \frac{1}{m}\nabla
S^{\hbar}(\textbf{x},t)$.

The convergence study of the determinist semi-classical case is
mathematically very difficult. We only study the example of a
coherent state where an explicit calculation is possible.

For the two dimensional harmonic oscillator,
$V(\textbf{x})=\frac{1}{2}m \omega^{2}\textbf{x}^{2}$, coherent
states are built~\cite{CohenTannoudji_1977} from the initial wave
function $\Psi_{0}(\textbf{x})$ which corresponds to the density
and initial action $ \rho^{\hbar}_{0}(\mathbf{x})= ( 2\pi \sigma
_{\hbar}^{2}) ^{-1} e^{-\frac{( \textbf{x}-\textbf{x}_{0})
^{2}}{2\sigma _{\hbar}^{2}}}$ and $
S_{0}(\mathbf{x})=S^{\hbar}_{0}(\mathbf{x})= m \textbf{v}_{0}\cdot
\textbf{x}$ with $ \sigma_\hbar=\sqrt{\frac{\hbar}{2 m \omega}}$.
Here, $\textbf{v}_0$ and $\textbf{x}_0$ are still constant vectors
and independent from $\hbar$, but $\sigma_\hbar$ will tend to $0$
as $\hbar$. With initial conditions, the density
$\rho^{\hbar}(\textbf{x},t)$ and the action
$S^{\hbar}(\textbf{x},t)$, solutions to the Madelung equations
(\ref{eq:Madelung1})(\ref{eq:Madelung2})(\ref{eq:Madelung3}), are
equal to ~\cite{CohenTannoudji_1977}:
$\rho^{\hbar}(\textbf{x},t)=\left( 2\pi \sigma_{\hbar} ^{2}
\right) ^{-1}e^{- \frac{( \textbf{x}-\xi(t)) ^{2}}{2\sigma_{\hbar}
^{2} }}$ and $S^{\hbar}(\textbf{x},t)= + m \frac{d\xi
(t)}{dt}\cdot \textbf{x} + g(t) - \hbar\omega t$, where $\xi(t)$
is the trajectory of a classical particle evolving in the
potential $V(\textbf{x})=\frac{1}{2} m \omega^{2} \textbf{x}^2 $,
with $\textbf{x}_0$ and $\textbf{v}_0$ as initial position and
velocity and $g(t)=\int _0 ^t ( -\frac{1}{2} m (\frac{d\xi
(s)}{ds})^{2} + \frac{1}{2} m \omega^{2} \xi(s)^2) ds$.
\begin{Theoreme}\label{t-convergenceparticulediscerne}~\cite{Gondran2011,Gondran2012}- When $\hbar\to 0$,
the density $\rho^{\hbar}(\textbf{x},t)$ and the action
$S^{\hbar}(\textbf{x},t)$ converge to
\begin{equation}
\rho(\textbf{x},t)=\delta( \textbf{x}- \xi(t)) ~~ and ~~
S(\textbf{x},t)= m \frac{d\xi (t)}{dt}\cdot\textbf{x} + g(t)
\end{equation}
where $S(\textbf{x},t)$ and the trajectory $\xi(t)$ are solutions
to the determinist Hamilton-Jacobi equations:
\begin{equation}\label{eq:statHJponctuelle1b}
\frac{\partial S\left(\textbf{x},t\right) }{\partial
t}|_{\textbf{x}=\xi(t)}+\frac{1}{2m}(\nabla S(\textbf{x},t)
)^{2}|_{\textbf{x}=\xi(t)}+V(\textbf{x})|_{\textbf{x}=\xi(t)}=0
\end{equation}
\begin{equation}\label{eq:statHJponctuelle1c}
\frac{d\xi(t)}{dt}=\frac{\nabla S(\xi(t),t)}{m}
\end{equation}
\begin{equation}\label{eq:statHJponctuelle1d}
S(\textbf{x},0)= m \textbf{v}_0 \cdot \textbf{x}
~~~and~~~\xi(0)=\textbf{x}_0.
\end{equation}
\end{Theoreme}
Therefore, the kinematic of the wave packet converges to the
single harmonic oscillator described by $\xi(t)$. Because this
classical particle is completely defined by its initial conditions
$\textbf{x}_0$ and $\textbf{v}_0$, it can be considered as \emph{a
discerned particle}. It is then possible to consider, unlike in
the statistical semi-classical case, that the wave function can be
viewed as a single quantum particle. The \textit{determinist
semi-classical case} is in line with the Copenhagen interpretation
of the wave function, which contains all the information on the
particle. A natural interpretation is proposed by
Schr\"odinger~\cite{Schrodinger_26} in 1926 for the coherent
states of the harmonic oscillator: the quantum particle is a
spatially extended particle, represented by a wave packet whose
center follows the classical trajectory.

\section{The non semi-classical case}

The Broglie-Bohm and Schrödinger interpretations correspond to the
semi-classical approximation.  They correspond to the two
interpretations proposed in 1927 at the Solvay congress by de
Broglie and Schrödinger. The principle of an interpretation that
depends on the particle preparation conditions is not really new.
It had already been figured out by Einstein and de Broglie. For
Louis de Broglie, its real interpretation was the double solution
theory introduced in 1927 in which the pilot-wave is just a
low-level product~\cite{Broglie}:

 "\textit{I introduced as a 'double solution theory' the idea that
it was necessary to distinguish two different solutions but both
linked to the wave equation, one that I called wave $u$ which was
a real physical wave but not normalizable having a local anomaly
defining the particle and represented by a singularity, the other
one as the Schr\"odinger $\Psi$ wave, which is normalizable
without singularities and being a probability representation.}"

We consider as interesting L. de Broglie's idea of the existence
of a statistical wave, $\Psi$ and of a soliton wave $u$; however,
it is not a double solution that appears here but a double
interpretation of the wave function according to the initial
conditions.

Einstein's point of view is well summed up in one of his final
papers (1953), "\textit{Elementary reflections concerning the
foundation of quantum mechanics}" in homage to Max
Born~\cite{Einstein}:

"\textit{The fact that the Schr\"odinger equation associated with
the Born interpretation does not lead to a description of the
"real states" of an individual system, naturally incites one to
find a theory that is not subject to this limitation.} \textit{Up
to now, the two attempts have in common that they conserve the
Schr\"odinger equation and abandon the Born interpretation.}
\emph{The first one, which marks de Broglie's comeback, was
continued by Bohm.... The second one, which aimed to get a "real
description" of an individual system and which might be based on
the Schr\"odinger equation is very late and is from Schr\"odinger
himself. The general idea is briefly the following : the function
$\psi$ represents in itself the reality and it is not necessary to
add it to Born's statistical interpretation.}[...] \textit{From
previous considerations, it results that the only acceptable
interpretation of the Schr\"odinger equation is the statistical
interpretation given by Born. Nevertheless, this interpretation
doesn't give the "real description" of an individual system, it
just gives statistical statements of entire systems.}"

Thus, it is because de Broglie and Schr\"odinger maintain the
Schrödinger equation that Einstein, who considers it as
fundamentaly statistical, rejected each of their interpretations.
Einstein thought that it was not possible to obtain an individual
deterministic behavior from the Schrödinger equation. It is the
same for Heisenberg who developped matrix mechanics and the second
quantization from the example of transitions in a hydrogen atom.

But there exist situations in which the semi-classical
approximation is not valid. It is in particular the case of state
transitions in a hydrogen atom. Indeed, since
Delmelt'experiment~\cite{Delmelt_1986} in 1986, the physical
reality of individual quantum jumps has been fully validated. The
semi-classical approximation, where the interaction with the
potential field can be described classically, is no longer
possible and it is necessary to quantify the electromagnetic field
since the exchanges occur photon by photon. In this situation, the
Schrödinger equation cannot give a deterministic interpretation
and the statistical Born interpretation seems to be the only valid
one. It was the third interpretation proposed in 1927 at the
Solvay congress, the interpretation that was recognized as the
right one in spite of Einstein's, de Broglie's and Schrödinger's
criticisms.

This doesn't mean that it is necessary to abandon determinism and
realism in quantum mechanics, but rather that the Schr\"odinger
wave function doesn't allow, in this case, to obtain an individual
behavior of a particle. An individual interpretation needs to use
the creation and annihilation operators of the quantum Field
Theory, but this interpretation still remains statistical.

We hypothesize that it is possible to construct a deterministic
quantum field theory that extends to the non semi-classical
interpretation of the double semi-classical interpretation. First,
as shown by de Muynck~\cite{Muynck}, we can construct a theory
with discerned (labeled) creation and annihilation operators in
addition to the usual non-discerned creation and annihilation
operators. But, to satisfy the determinism, it is necessary to
search, at lower scale, the mechanisms that allow the emergence of
the creation operator.

\section{Conclusion}

The study of the convergence of the Madelung equations when $
h\rightarrow 0$, gives the following results:

- In the statistical semi-classical case, the quantum particles
converge to classical non-discerned ones satisfying the
statistical Hamilton-Jacobi equations, and the Broglie-Bohm
pilot-wave interpretation is relevant.

- In the determinist semi-classical case, the quantum particles
converge to classical discerned ones satisfying the determinist
Hamilton-Jacobi equations. And we can make a realistic and
deterministic assumption such as the Schr\"odinger interpretation.

This double interpretation seems to be the interpretation of Louis
de Broglie's "double solution" idea.

- In the case where the semi-classical approximation is no longer
valid, as in the transition states in the hydrogen atom, Louis de
Broglie's "double solution" is not directly applicable. But, we
hypothesize that it is possible to construct a deterministic
quantum field theory that extends this double interpretation to
the non semi-classical case.

\end{document}